\documentclass[%
 reprint,
 amsmath,amssymb,
 aps,
]{revtex4-2}

\usepackage[dvipsnames]{xcolor}
\usepackage{graphicx}
\usepackage{placeins}
\usepackage{dcolumn}
\usepackage{bm}

\begin{document}

\preprint{APS/123-QED}

\title{Dynamics of reconfigurable straw-like elements}

\author{Dotan Ilssar$^{1,2}$}
\email{dilssar@ethz.ch. Author to whom correspondence should be addressed.}
\author{Michael Pukshansky$^{1}$}
\author{Yizhar Or$^{1}$}
\author{Amir D. Gat$^{1}$}
\email{amirgat@technion.ac.il}
\affiliation{$^{1}$Faculty of Mechanical Engineering, Technion – Israel Institute of Technology, Haifa 3200003, Israel\\ $^{2}$Mechanics \& Materials Lab, Department of Mechanical and Process Engineering, ETH Zürich, Zürich 8092, Switzerland}

\date{\today}

\begin{abstract}
In this paper, we discuss the dynamic modeling of fluid-filled straw-like elements consisting of serially interconnected elastic frusta with both axisymmetric and antisymmetric degrees of freedom, assuming planar motion. Under appropriate conditions each sub-structure has four stable equilibrium states. This gives the system under investigation the ability to remain stable in a large number of complex states, which is a vital ability for myriad of applications, including reconfigurable structures and soft robots. The theoretical model explains the dynamics of a single straw-like element in a discrete manner, considering inertial, damping, and gravitational effects, while taking into account the nonlinear elasticity of the elastic frusta, and assuming hydrostatic behavior of the entrapped fluid. After identifying the geometric and elastic parameters of the theoretical model based on relatively simple experiments, the model is validated compared to numerical simulations and experiments. The numerical simulations validate the theoretical elasticity of the elastic frusta, whereas the overall dynamic behavior of the system and the influence of unmodeled fluidic effects are examined experimentally. It is demonstrated both theoretically and empirically that straw-like elements cannot be adequately modeled using simple uniaxial deformations. In addition, the experimental validation indicates that the suggested model can accurately capture their overall dynamics.
\end{abstract}


\maketitle

\section{Introduction}
Reconfigurable metamaterials capable of changing their forms and mechanical properties are an emerging field in mechanics and soft robotics. Recently, instability-based reconfigurable metamaterials have gained great importance thanks to their unique characteristics and performance \cite{bertoldi2017flexible}. For example, hierarchical structures composed of repeated multistable elements in different configurations show extreme properties such as large zero Poisson's ratio deformations in the elastic regime \cite{yang2019multi,hua2019multistable,findeisen2017characteristics}, as well as multiaxial complex stable states \cite{zhang2020design}. The various stable equilibria of instability-based metamaterials also lead to exotic dynamic behaviors, allowing for instance to alter the propagation characteristics of elastic waves \cite{meaud2017tuning}. Another important dynamic property achievable in arrays of multistable elements is energy localization. Indeed, recent studies demonstrated the creation of propagating wave fronts along one- and two-dimensional arrays of bistable elements \cite{raney2016stable,nadkarni2016unidirectional,khajehtourian2020phase,jin2020guided,katz2018solitary}. Finally, the ability to employ a single input to bring a configuration of discretely \cite{ben2020single,glozman2010self} and continuously \cite{peretz2020underactuated} interconnected bistable elements from one stable equilibrium state to another, was achieved by embedding these locally bistable structures with viscous fluid. By these means, a change of pressure at the inlet creates a pressure wave that folds or deploys different sections of the structure in order, allowing it to be brought to any designated stable state, while passing through undesired configurations.

The present paper deals with modeling and investigating the dynamic behavior of locally multistable mechanical elements inspired by ‘bendy straws’, to pave the way towards inflatable reconfigurable structures utilizing these elements as building blocks. Thanks to their large number of stable equilibria, structures consisting of such straw-like elements can be brought to myriad of complex operative configurations, by governing the pressure of the fluid trapped inside them. A recent study started shedding light on the static behavior of straw-like elements by showing empirically and theoretically that their multistability  originates in internal stresses \cite{bende2018overcurvature}. A similar local multistability was demonstrated in an origami implementation of ‘bendy straws’, where pop-through defects gave rise to a single non-axial stable state of each unit-cell \cite{bernardes2022design}. In another recent article, the authors produced a two dimensional array of uncoupled straws restricted to deform uniaxially \cite{pan20193d}. This configuration has an unprecedented number of stable states which provide a high flexibility while determining its reconfigurable form and local stiffness. Nevertheless, no available model describing the static and dynamic behaviors of straw-like elements was found in the literature. Such model can enable to theoretically design the properties and operating conditions of these elements, to achieve designated behaviors.

Indeed, here we start with theoretical modeling of a single straw-like element consisting of serially connected elastic conical frusta, whose dynamics is governed by an externally applied pressure at a reservoir supplying incompressible fluid through a slender channel. Considering a degenerated two-dimensional representation of the system, each constituent frustum has both axisymmetric and antisymmetric degrees of freedom (DOFs), where due to a nonlinear elastic behavior, under appropriate geometrical and loading conditions each frustum has four distinct stable equilibria. The proposed theoretical model considers inertial and gravitational effects assuming a nearly uniform pressure field, as well as a simplified description of the dissipative forces accounting for both fluidic effects and structural damping. The model also considers the nonlinear stiffness of the elastic frusta utilizing an extension of the well-known formulation by Almen and Laszlo \cite{almen1936uniform}, and the fluidic effects related to the flow inside the channel which is assumed laminar at high Reynolds numbers. Following the theoretical derivation, we introduce an efficient system identification process based on measurements and relatively simple experiments, used to determine the geometrical parameters of the system, as well as the physical parameters related to the solid. This process is utilized to calibrate the theoretical model so it captures the behavior of an experimental demonstrator that was designed and manufactured for model verification. The nonlinear quasi-static theoretical behavior of a single elastic frustum is then investigated and validated compared to finite element simulations. These simulations are further exploited as a basis for a numerical justification to disregard the deformations of half of the constituent frusta, which significantly reduces the number of DOFs. Finally, the experimental demonstrator is utilized for verifying the overall dynamic behavior of the theoretical model, and to assess unmodeled fluidic effects by calibrating the parameters which are affected by them. The highly correlated analytical and numerically calculated quasi-static behaviors of the system predict that when the inertial effects are weak, each frustum should pass through a partially-snapped state, while switching between fully-snapped configurations. This claim is experimentally confirmed as under negative external gauge pressure, the different sections of the demonstrator are folded according to this theoretically predicted response. Nevertheless, it is shown that while deploying the system by applying positive external gauge pressure, due to lower damping reflected in high inertial forces, each segment can be immediately brought from one fully-snapped state to another.

\section{Model derivation} \label{sec2}
\begin{figure}
  \includegraphics[width=\linewidth]{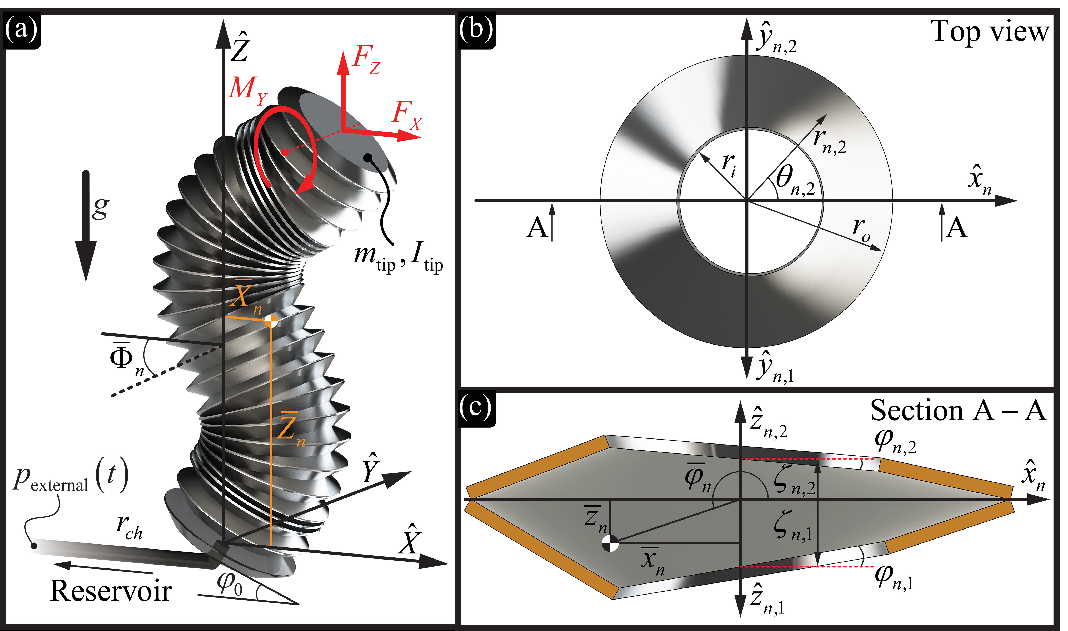}
  \caption{(a) Schematic layout of the system under investigation, consisting of a liquid-filled straw-like element which is connected through a slender channel to a liquid reservoir with externally dictated static pressure. (b, c) Top and section views of a constituent multistable cell, composed of two elastic conical frusta.}
  \label{Figure1}
\end{figure}

The system under investigation consists of a liquid-filled straw which is sealed on one end, where its second end is connected to a reservoir whose pressure is externally dictated, see Figure \ref{Figure1} (a). Here, we present the formulation of a simplified model capturing the governing dynamics of this system, starting with descriptions of the inertial and dissipative effects as well as the influence of external and gravitational forces. Next, we supplement the model with the strain energy function of the system, derived based on a quasi-static analysis of a single elastic frustum, which is the basic building block of the system.

\subsection{Formulation of the dynamic and external effects}

As illustrated in Figure \ref{Figure1}, we describe the straw as a serial interconnection of $N$ liquid-filled cells, each consisting of two elastic frusta. For simplicity we consider a two-dimensional problem where the motion of the system and its constituent building blocks are bounded to the $\hat X-\hat Z$ plane of the global coordinate system whose origin is located at the middle of the straw’s base, see Figure \ref{Figure1} (a). Under this simplification, the deformation of each multistable cell is formulated in terms of ${\zeta _{n,1}}$ and ${\varphi _{n,1}}$ denoting the axisymmetric and antisymmetric DOFs of the frustum closer to the inlet, as well as ${\zeta _{n,2}}$ and ${\varphi _{n,2}}$  describing the corresponding DOFs of the frustum that is farther from the inlet. Namely, under the assumption of small angles, the axial position of each material point in the k\textsuperscript{th} frustum of the n\textsuperscript{th} multistable cell, with respect to the cell’s middle surface, is 
\begin{equation}\label{eq1}
{z_{n,k}}\left( {{r_{n,k}},\,{\theta _{n,k}}} \right) \approx \left( {{\zeta _{n,k}} - {r_i}{\varphi _{n,k}}\cos {\theta _{n,k}}} \right)\frac{{{r_o} - {r_{n,k}}}}{{{r_o} - {r_i}}}.
 \end{equation}
 Here, ${r_{n,k}} \in \left[ {{r_i},\,{r_o}} \right]$ and ${\theta _{n,k}} \in \left[ {0,\,2\pi } \right)$ are the local radial and tangential coordinates of the relevant frustum, where $r_i$ and $r_o$ respectively describe the inner and outer radii of all frusta. The representation in (\ref{eq1}) neglects radial deformations, thus keeps the radial cross-sections of the frusta as well as their bases, undistorted. Next, the channel connecting the straw to the liquid reservoir whose dictated time-dependent static pressure is denoted ${p_{{\rm{external}}}}\left( t \right)$, is considered to be slender compared to both the straw and the reservoir, where its radius is denoted $\,{r_{ch}}\,$. Thus, referring to the fluid in both the straw and the reservoir as semi-infinite media, their pressure fields are approximated uniform, whereas assuming incompressible laminar flow at high Reynolds numbers, the fluid inside the channel is described by the steady Bernoulli equation. It was previously shown that under these conditions, the flow regimes throughout filling and depletion of a closed vessel are radically asymmetric due to boundary layer separation, giving rise to an internal jet when the fluid flows into the vessel \cite{ilssar2020inflation}. In this case, the flow velocities in the vessel being filled are significantly higher than those at the other end of the channel filling it, where the dynamic pressure can be neglected. Therefore, while inflating the straw, the dictated pressure is referred to as the stagnation pressure of the closed system. However, during deflation, boundary layer separation occurs in the liquid reservoir, thus the stagnation pressure is taken as the uniform pressure inside the straw. Consequently, denoting the density of the fluid by $\rho$, the uniform pressure inside the straw is taken as 
\begin{equation}\label{eq2}
p\left( t \right) = {p_{{\rm{external}}}}\left( t \right) - \frac{\rho }{{2{\pi ^2}r_{ch}^4}}{\left( {\sum\limits_{n = 1}^N {\frac{{{\rm{d}}{V_n}}}{{{\rm{d}}t}}} } \right)^2}{\rm{sgn}}\left( {\sum\limits_{n = 1}^N {\frac{{{\rm{d}}{V_n}}}{{{\rm{d}}t}}} } \right),
\end{equation}
 where $V_n$ is the volume of the fluid residing inside the n\textsuperscript{th} multistable cell. As discussed above, for simplicity the radial deformations of the different frusta are assumed to be negligible, meaning that their radial dependency is represented by the linear function given in (\ref{eq1}). Integration over the volume of the n\textsuperscript{th} multistable cell utilizing this assumption yields the following expression, describing in terms of the generalized coordinates ${\zeta _{n,k}}$ and ${\varphi _{n,k}}$, the volume of the fluid enclosed within this cell: \[V_n=\pi \frac{\left(r_o^2 + r_i^2 + r_o r_i \right) \left(\zeta _{n,1}+\zeta _{n,2}\right)}{3}.\]
 
 Assuming that the inertial and body forces of the solid are negligible compared to those of the entrapped fluid due to low thickness of the straw, the effects of the solid's mass are disregarded. However, for practical reasons such as the need of a relatively heavy body to seal the tip of the straw, the mass and moment of inertia of this sealing denoted \({m_{{\rm{tip}}}}\) and \({I_{{\rm{tip}}}}\) are taken into account. Further, in agreement with the assumption claiming that the pressure field inside the straw is uniform, the motion of the fluid trapped in each cell is treated as a quasi-rigid body, whose instantaneous mass and moment of inertia are determined by the DOFs of its constituent frusta. Thus, the contribution of each multistable cell to the total kinetic energy is computed considering the translation of its center of mass and the rotation around this point, considering its varying volume and shape. Summing the kinetic energies of the sealing as well as those related to all cells leads to the following expression, describing the overall kinetic energy of the system:
\begin{equation}\label{eq3}
\begin{array}{l}
{\cal T} = \sum\limits_{n = 1}^N {\left\{ {\frac{{\rho {V_n}}}{2}\left[ {{{\left( {\frac{{{\rm{d}}{{\bar X}_n}}}{{{\rm{d}}t}}} \right)}^2} + {{\left( {\frac{{{\rm{d}}{{\bar Z}_n}}}{{{\rm{d}}t}}} \right)}^2}} \right] + \frac{{{I_n}}}{2}{{\left[ {\frac{{{\rm{d}}\left( {{{\bar \Phi }_n} + {{\bar \varphi }_n}} \right)}}{{{\rm{d}}t}}} \right]}^2}} \right\}} \\
\,\,\,\,\,\,\,\,\,\,\,\,\,\,\,\,\,\,\,\,\, + \frac{{{m_{{\rm{tip}}}}}}{2}\left[ {{{\left( {\frac{{{\rm{d}}{{\bar X}_{{\rm{tip}}}}}}{{{\rm{d}}t}}} \right)}^2} + {{\left( {\frac{{{\rm{d}}{{\bar Z}_{{\rm{tip}}}}}}{{{\rm{d}}t}}} \right)}^2}} \right] + \frac{{{I_{{\rm{tip}}}}}}{2}{\left( {\frac{{{\rm{d}}{{\bar \Phi }_{{\rm{tip}}}}}}{{{\rm{d}}t}}} \right)^2}.
\end{array}
\end{equation}
Here, \[{\bar X_n} = \sum\limits_{j = 1}^{n - 1} {\left( {{\zeta _{j,1}} + {\zeta _{j,2}}} \right)\sin {{\bar \Phi }_j}}  + \left( {{\zeta _{n,1}} + {{\bar z}_n}} \right)\sin {\bar \Phi _n} + {\bar x_n}\cos {\bar \Phi _n},\] and \[{\bar Z_n} = \sum\limits_{j = 1}^{n - 1} {\left( {{\zeta _{j,1}} + {\zeta _{j,2}}} \right)\cos {{\bar \Phi }_j}}  + \left( {{\zeta _{n,1}} + {{\bar z}_n}} \right)\cos {\bar \Phi _n} - {\bar x_n}\sin {\bar \Phi _n},\] are the horizontal and vertical coordinates of the n\textsuperscript{th} multistable cell's center of mass, with respect to the global coordinate system. Furthermore, \[{\bar \Phi _n} = {\varphi _0} + \sum\limits_{j = 1}^{n - 1} {\left( {{\varphi _{j,1}} + {\varphi _{j,2}}} \right)}  + {\varphi _{n,1}},\] is the orientation of the cell's middle surface around the $\hat Y$ axis, where ${\varphi _0}$ is the inclination angle of the straw's base, relative to the horizon. Similarly, \[{X_{{\rm{tip}}}} = \sum\limits_{j = 1}^N {\left( {{\zeta _{j,1}} + {\zeta _{j,2}}} \right)\sin {{\bar \Phi }_j}},\] \[{Z_{{\rm{tip}}}} = \sum\limits_{j = 1}^N {\left( {{\zeta _{j,1}} + {\zeta _{j,2}}} \right)\cos {{\bar \Phi }_j}},\] and \[{\Phi _{{\rm{tip}}}} = {\varphi _0} + \sum\limits_{j = 1}^N {\left( {{\varphi _{j,1}} + {\varphi _{j,2}}} \right)}, \] represent the horizontal, vertical and rotational coordinates of the straw's sealed end. Next, $\bar x_n$, $\bar z_n$ and $\bar \varphi_n$ are the translational and angular coordinates of the n\textsuperscript{th} multistable cell's center of mass, with respect to the local coordinate system of its constituent frustum which is farther from the straw's base, see Figure \ref{Figure1} (c). Assuming ${\varphi _{n,1}},\,{\varphi _{n,2}} \ll 1$ and that the frusta do not deform radially, these local coordinates are given by the following approximated expressions: \[ \bar x_n=-r_o r_i \frac{\varphi_{n,1}+\varphi_{n,2}}{4 \left(\zeta _{n,1} + \zeta _{n,2}\right)}, \] \[\begin{array}{l}{\bar z_n} \approx \frac{{2\left( {r_o^2 + 3r_i^2 + 2{r_o}{r_i}} \right)\left( {\zeta _{n,2}^2 - \zeta _{n,1}^2} \right) + {r_o}r_i^2\left( {{r_o} + 2{r_i}} \right)\left( {\varphi _{n,2}^2 - \varphi _{n,1}^2} \right)}}{{8\left( {r_o^2 + r_i^2 + {r_o}{r_i}} \right)\left( {{\zeta _{n,1}} + {\zeta _{n,2}}} \right)}}\end{array},\] and \({\bar \varphi _n} = {\rm{atan}}2\left( {{{\bar z}_n},{{\bar x}_n}} \right)\).
The last component in the expression of the kinetic energy to be derived is the moment of inertia of the n\textsuperscript{th} multistable cell around its center of mass. Under the underlying assumptions this expression is given by \[\begin{array}{l}
{I_n} \approx \pi \rho \frac{\left(r_o^2 + 6r_i^2 + 3 r_o  r_i \right) \left(\zeta _{n,1}^3 + \zeta _{n,2}^3 \right)}{30}\\ \,\,\,\,\,\,\,\,\,
+\pi \rho r_i^2 \frac{\left(r_o^2 + r_i^2 + 3 r_o r_i \right) \left(\zeta _{n,1}\varphi _{n,1}^2 + \zeta _{n,2} \varphi _{n,2}^2 \right)}{20}\\ \,\,\,\,\,\,\,\,\, +\pi \rho \frac{\left( r_o^5 - r_i^5 \right) \left(\zeta_{n,1} + \zeta_{n,2} \right)}{20 \left(r_o-r_i\right)} - \rho {V_n}\left( {\bar x_n^2 + \bar z_n^2} \right).
\end{array}\]

The Rayleigh dissipation function emanating from structural damping and the interaction between the solid and the entrapped and ambient fluids, is considered linearly dependent on the time derivatives of the DOFs, thus is taken as
\begin{equation}\label{eq4}
{\cal D} = \sum\limits_{n = 1}^N {\sum\limits_{k = 1}^2 {\left[ {\frac{{{c_\zeta }}}{2}{{\left( {\frac{{{\rm{d}}{\zeta _{n,k}}}}{{{\rm{d}}t}}} \right)}^2} + \frac{{{c_\varphi }}}{2}{{\left( {\frac{{{\rm{d}}{\varphi _{n,k}}}}{{{\rm{d}}t}}} \right)}^2}} \right]} }.
\end{equation}
Here, for simplicity the damping coefficients of all axisymmetric and antisymmetric DOFs, denoted ${c_\zeta }$ and ${c_\varphi }$, are considered identical.

Finally, we formulate the virtual work considering the uniform pressure applied to the straw's sealed end assuming the terminal radius equals to $r_i$, as well as the horizontal and vertical localized forces $F_X$ and $F_Z$ and bending moment $M_Y$, exerted on this end. Moreover, the gravitational forces acting on the sealing as well as on the fluid trapped inside the straw's cells are also taken into account. Considering all the effects mentioned above, the total virtual work is given by:
\begin{equation}\label{eq5}
\begin{array}{l}
\delta {\cal W} =  - \rho g\sum\limits_{n = 1}^N {{V_n}\delta {{\bar Z}_n}}  + \left( {{F_X} + \pi r_i^2p\sin {{\bar \Phi }_{{\rm{tip}}}}} \right)\delta {\bar X_{{\rm{tip}}}} \\ \,\,\,\,\,\,\,\,\,\,\,\,\,\,\,\,\,+ \left( {{F_Z} + \pi r_i^2p\cos {{\bar \Phi }_{{\rm{tip}}}} - {m_{{\rm{tip}}}}g} \right)\delta {\bar Z_{{\rm{tip}}}} \\ \,\,\,\,\,\,\,\,\,\,\,\,\,\,\,\,\,+ {M_Y}\delta {\bar \Phi _{{\rm{tip}}}},
\end{array}
\end{equation}
where $g$ is the gravitational acceleration and $\delta  \bullet $ is a variation in $ \bullet $.

\subsection{Adding the effects of the solid’s elasticity} \label{sec22}

To complete the formulation of the system's dynamics, it is necessary to derive the potential energy related to the elasticity of the frusta, and to the direct influence of the uniform pressure field on them. Thus, in this section a quasi-static analysis of a single elastic conical frustum is carried out, while omitting the subscript $n,k$ for brevity. This results in the potential energy of a general frustum, which is then used to describe all constituent frusta. The latter leads to the overall potential energy of the straw assuming the creases connecting the different frusta are extremely thin. Namely,  the contribution of these creases to the strain energy is negligible, as well as the tangential moments they apply to the bases of the frusta.

The analysis is based on an article by Almen and Laszlo \cite{almen1936uniform} who modelled the deformations of a Belleville washer undergoing axial loading, and some more recent studies that corrected and modified this model, yet still consider only axial deflections \cite{mastricola2017analytical,la2001stiffness}. Thus, the underlying assumptions of these papers are adopted. These assumptions claim that the frustum is thin, its material is linear, and that the deformations of its cross-sections are minor, meaning that the bending and elongation along the radial direction are disregarded. Instead, each radial cross-section is considered rotating around a neutral point, located on its natural axis at a radial location \({r_{{\rm{neutral}}}}\) which should be determined. Consequently, the radial and axial components of the stress are neglected. It should be noted that since the model derived here considers deformations that deviate from axisymmetry, \({r_{{\rm{neutral}}}}\) is initially considered dependent on the tangential coordinate $\theta$. The rotations of the cross-sections result in direct tangential bending stresses, and additional tensile stresses caused due to radial displacements which stretch each material point tangentially. These two stress components apply internal moments which balance those caused due to external loads, leading to the elastic behavior of the frustum.

\begin{figure}
  \includegraphics[width=\linewidth]{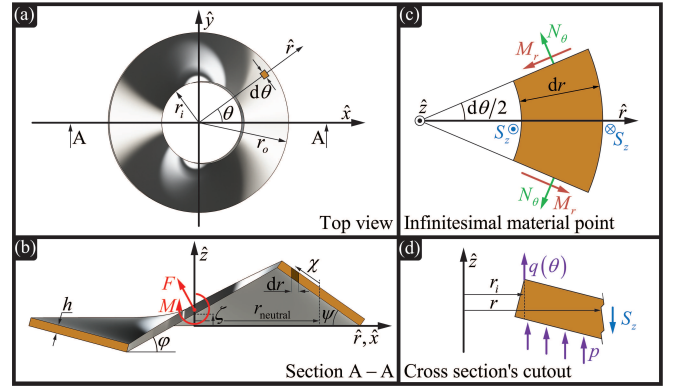}
  \caption{Schematic layout of an elastic conical frustum. (a, b) Top and section views of the entire frustum, (c) Top view of an arbitrary infinitesimal material point, (d) Cutout of an arbitrary cross-section.}
  \label{Figure2}
\end{figure}

The first contribution to the internal moment to be formulated is the one caused due to the tangential stretching of frustum. For this sake, the tangential strain of a general infinitesimal material point is calculated as function of the frustum's base angle taken as $\psi  = {\psi _0} + \Delta {\psi _1} + \Delta {\psi _2}\cos \theta $, see Figure \ref{Figure2} (b), where $\theta $ is the tangential coordinate revolving around the axis of the frustum denoted $\hat z$, see Figure \ref{Figure2} (a). This form allows the frustum to undergo both axisymmetric and antisymmetric deflections related to $\Delta {\psi _1}$ and $\Delta {\psi _2}$ respectively, from the stress-free axisymmetric state where the base angle is constant and equals to $\psi_0$. Employing the above notations and assuming small base angles such that $\psi  \ll 1$, the local tangential strain is given by \cite{almen1936uniform} \[\varepsilon_{\theta,1}=\frac{\left( r_{\rm{neutral}}-\chi \cos \psi \right){\rm{d}} \theta}{\left( r_{\rm{neutral}}-\chi \cos \psi_0 \right){\rm{d}} \theta}-1\approx \frac{\chi \left(\psi^2 - \psi_0^2 \right)}{2 \left(r_{\rm{neutral}} - \chi \right)},\] where $\chi \left(\theta \right) $ is a cross-section dependent coordinate which is tangent to the frustum's face, whose origin is located at ${r_{{\rm{neutral}}}}\left(\theta \right)$, see Figure \ref{Figure2} (b). Under the assumptions that the material is linear and that the radial and axial stress components are negligible, Hooke's law relates the tangential stress and strain by \({\sigma _{\theta,1}} = E{\varepsilon _{\theta,1} }\), where $E$ is the solid's Young modulus. Utilizing this stress component, the contribution of a general infinitesimal material point’s tangential stretching, to the internal radial moment per unit length around the cross-section's neutral point is given by \[{M_{r{\rm{,1}}}} = \left[ {\chi \sin \psi \hat z \times {\sigma _{\theta ,1}}h\left( { - \hat \theta } \right)} \right] \cdot \hat r \approx \frac{{Eh{\chi ^2}\psi \left( {{\psi ^2} - \psi _0^2} \right)}}{{2\left( {{r_{{\rm{neutral}}}} - \chi } \right)}},\] where $h$ is the thickness of the frustum.

As mentioned above, the second contribution to the internal radial moment per unit length is related to the tangential bending of the frustum. To formulate this contribution, first the base angle dependent local change of the tangential curvature is computed under the assumptions mentioned above, and is taken as \cite{timoshenko1956strength} \[\Delta {\kappa _\theta } = \frac{{\sin \psi }}{{{r_{{\rm{neutral}}}} - \chi \cos \psi }} - \frac{{\sin {\psi _0}}}{{{r_{{\rm{neutral}}}} - \chi \cos {\psi _0}}} \approx \frac{{\psi  - {\psi _0}}}{{{r_{{\rm{neutral}}}} - \chi }}.\] Next, the desired expression is computed by integrating the resultant tangential stress multiplied by $z$, over the thickness of the frustum \cite{reddy2017energy}. Considering the solid behaves according to Kirchhoff plate theory, and recalling that the axial and radial stresses are nulled, the tangential stress component is given by ${\sigma _{\theta ,2}} =  - Ez\Delta {\kappa _\theta }$, thus the corresponding radial moment per unit length is \({M_{r{\rm{,2}}}} \approx {{E{h^3}\Delta {\kappa _\theta }} \mathord{\left/
 {\vphantom {{E{h^3}\Delta {\kappa _\theta }} {12}}} \right.
 \kern-\nulldelimiterspace} {12}}\). Summation of \({M_{r{\rm{,1}}}}\) and \({M_{r{\rm{,2}}}}\), yields the overall internal radial moment per unit length.

The next stage is calculating the radial positions of the cross-sections’ neutral points. This is done by balancing the internal normal forces $N_\theta$ acting on the radial faces of an arbitrary infinitesimal sector due to the overall tangential stress $\sigma_{\theta}$ computed above. Namely, the normal forces which are shown in Figure \ref{Figure2} (c) for an infinitesimal material point, are integrated over the two faces of the sector, and projected on the radial direction. From static considerations, these two terms should cancel each other, which leads to an expression which indicates that in every cross-section the neutral point resides in \({r_{{\rm{neutral}}}} \approx {{\left( {{r_o} - {r_i}} \right)} \mathord{\left/
 {\vphantom {{\left( {{r_o} - {r_i}} \right)} {\ln \left( {{{{r_o}} \mathord{\left/
 {\vphantom {{{r_o}} {{r_i}}}} \right.
 \kern-\nulldelimiterspace} {{r_i}}}} \right)}}} \right.
 \kern-\nulldelimiterspace} {\ln \left( {{{{r_o}} \mathord{\left/
 {\vphantom {{{r_o}} {{r_i}}}} \right.
 \kern-\nulldelimiterspace} {{r_i}}}} \right)}}\).

To balance the internal moment formulated above, the influence of the external loading applied to the frustum is derived. This is done while considering a distributed load given by \(q\left( \theta  \right) = \left( {{F \mathord{\left/
 {\vphantom {F {2\pi {r_i}}}} \right.
 \kern-\nulldelimiterspace} {2\pi {r_i}}}} \right) - \left( {{M \mathord{\left/
 {\vphantom {M {\pi r_i^2}}} \right.
 \kern-\nulldelimiterspace} {\pi r_i^2}}} \right)\cos \theta \), representing the interaction with the neighboring frusta which introduce both normal force and bending moment denoted $F$ and $M$, to the frustum’s edges. Moreover, the direct influence of the entrapped fluid's uniform pressure is also taken into account. Consequently, under the assumption of small angles enabling to refer to both the pressure and the applied loads as purely uniaxial, the resulting shear force per unit length acting on an arbitrary radial cutout is given by \[{S_z} \approx \frac{{2{r_i}q\left( \theta  \right){\rm{ + }}\left[ {{{\left( {{r_{{\rm{neutral}}}} - \chi } \right)}^2} - r_i^2} \right]p}}{{2\left( {{r_{{\rm{neutral}}}} - \chi } \right)}},\] see Figure \ref{Figure2} (d). Finally, a moment balance considering the internal moments as well as the shear force couple acting on the general infinitesimal material point in Figure \ref{Figure2} (c), is applied after projection on the axis perpendicular to the radial coordinate. Integration of the expression obtained by these means along the radial dimension of the frustum, and elimination of small terms which are dependent on the differential radius and angle ${\rm{d}}r$ and $\rm{d} \theta$ leads to the quasi-static behavior of the frustum. The latter is then converted to terms of the axisymmetric and antisymmetric DOFs by their relations with the base angles given by \(\zeta  \approx \left( {{r_o} - {r_i}} \right)\left( {{\psi _0} + \Delta {\psi _1}} \right)\) and \(\varphi  \approx {{\left( {{r_i} - {r_o}} \right)\Delta {\psi _2}} \mathord{\left/
 {\vphantom {{\left( {{r_i} - {r_o}} \right)\Delta {\psi _2}} {{r_i}}}} \right.
 \kern-\nulldelimiterspace} {{r_i}}}\), see Figure \ref{Figure2} (b). The resulting formulation  consists of terms dependent on $\cos j\theta $ where $j = 0, \cdots ,3$, which can be separated into four equations thanks to orthonormality. Here, the equations related to the terms multiplying $\cos 2\theta$ and $\cos 3\theta$ can be balanced by considering higher harmonics of the distributed load ${q\left( \theta  \right)}$ but are disregarded due to irrelevance. Yet, the other two equations which are related to the leading harmonics, yield the following closed form expressions of the axial force and bending moment applied externally by the neighboring frusta:
 \begin{subequations}
 \begin{equation}\label{eq6a}
F \approx 4{C_1}\zeta \left( {2{\zeta ^2} - 2\zeta _0^2 + 3r_i^2{\varphi ^2}} \right) + 2{C_2}\left( {\zeta  - {\zeta _0}} \right) - {C_3}p,
\end{equation}
 \begin{equation}\label{eq6b}
M \approx r_i^2\varphi \left[ {{C_1}\left( {12{\zeta ^2} - 4\zeta _0^2 + 3r_i^2{\varphi ^2}} \right) + {C_2}} \right],
\end{equation}
\end{subequations}
where \({\zeta _0} \approx \left( {{r_o} - {r_i}} \right){\psi _0}\) is the load-free deflection of the frustum, whereas \[{C_1} = \frac{\pi E h}{8 \left( r_o - r_i \right)^3} \left(\frac{r_o+r_i}{2}-\frac{r_o-r_i}{\ln \left(r_o/r_i\right)}\right),\] \[{C_2} = \frac{\pi E h^3 \ln \left(r_o/r_i \right)}{12 \left(r_o-r_i \right)^2}, \]  and \[{C_3} = \pi \frac{r_o^2-2 r_i^2+r_o r_i}{3},\] are constant coefficients.
 
 Combining the integrations of equations (\ref{eq6a}) and (\ref{eq6b}) with respect to the axisymmetric and antisymmetric DOFs, followed by retrieving the general subscript $n,k$ yields the potential energy of a general frustum, given by:
  \begin{equation}\label{eq7}
\begin{array}{l}
{{\cal V}_{n,k}} \approx 2{C_1}\zeta _{n,k}^4 + \left( {{C_2} - 4{C_1}\zeta _{0,k}^2} \right)\zeta _{n,k}^2 \\ \,\,\,\,\,\,\,\,\,\,\,\,\,\,\,\,\,- \left( {2{C_2}{\zeta _{0,k}} + {C_3}p} \right){\zeta _{n,k}} + \frac{{3{C_1}r_i^4}}{4}\varphi _{n,k}^4 \\ \,\,\,\,\,\,\,\,\,\,\,\,\,\,\,\,\,+ \frac{{{C_2} - 4{C_1}\zeta _{0,k}^2}}{2}r_i^2\varphi _{n,k}^2 + 6{C_1}r_i^2\zeta _{n,k}^2\varphi _{n,k}^2 + Const.
\end{array}
\end{equation}
Here, assuming all odd frusta are identical as well as all even frusta, ${\zeta _{0,k}}$ denotes the load-free deflection of the k\textsuperscript{th} frusta in each n\textsuperscript{th} multistable cell. 
The overall potential energy of the system in Figure \ref{Figure1} (a) is computed by the summation of those corresponding to all frusta, thus is given by \({\cal V} = \sum\limits_{n = 1}^N {\sum\limits_{k = 1}^2 {{{\cal V}_{n,k}}} } \). Therefore, applying the Hamilton’s principle (e.g. \cite{geradin2014mechanical}) on this summation alongside the kinetic energy, dissipation function, and virtual work derived in the previous section yields a system of $4N$ ordinary differential equations, governing the dynamics of the system under investigation.

\section{Results} \label{sec3}

\begin{figure}
   \begin{center}
  \includegraphics[width=\linewidth]{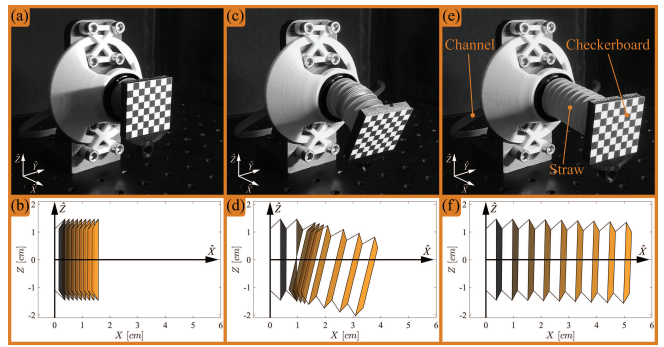}
  \end{center}
  \caption{(a,c,e) The experimental demonstrator in three zero gauge pressure, statically stable states. (b,d,f) The corresponding theoretically predicted states.}
  \label{Figure3}
\end{figure}

The theoretical model derived above is examined based on the experimental demonstrator presented in Figure \ref{Figure3}. This figure displays three of the demonstrator's stable states alongside their corresponding theoretical forms achieved numerically from the model, considering body forces where the gravitational acceleration is taken as $g = {\rm{9.795}}$ ${{\rm{m}} \mathord{\left/
 {\vphantom {{\rm{m}} {{{\rm{s}}^{\rm{2}}}}}} \right.
 \kern-\nulldelimiterspace} {{{\rm{s}}^{\rm{2}}}}}$. The demonstrator consists of a water-filled truncated off-the-shelf toy straw with nine multistable cells, clamped horizontally such that $\varphi_0=90$ $\rm{deg}$. In agreement with the theoretical system, the fixed end of the straw is connected to a water reservoir through a narrow channel, whereas its free end is sealed with a 3D printed part. The estimated mass and moment of inertia of this sealing with respect to the center of the face which is farthest from the inlet are taken as ${{m_{{\rm{tip}}}}}=13.45$ gr and  ${{I_{{\rm{tip}}}}}=2986.91$ ${\rm{gr}} \cdot {{\rm{mm}}^2}$, considering the entrapped fluid. The motion of the demonstrator is governed by an ELVEFLOW OB1 MK3 piezoelectric pressure controller having a settling time of down to 35 ms, connected to the reservoir through a second channel which is filled with air. The two channels as well as the water reservoir and the connections between the different elements constitute a complex channel system whose effective geometry is to be determined empirically, see section \ref{sec31}. Finally, the motion of the system is captured by tracking a ${38.5\times38.5}$ ${\rm{mm}^2}$ checkerboard with an ${8\times8}$ grid glued to the free end of the straw, utilizing a GoPro HERO6 BLACK camera, capturing 240 frames per second in a resolution of 1080p. Here, since there is an offset of approximately 8.8 mm between the checkerboard's plane and the face of the ninth multistable cell which is considered the sealed end, this offset is subtracted from the measurements in post processing.

To examine the theoretically predicted behavior of the system based on the experimental demonstrator, the model's physical and geometrical parameters should be computed. Indeed, the upcoming section presents the algorithm used to efficiently estimate most parameters of the system, alongside a numerically justified assumption made in order to simplify the model, for reducing the computational effort of its simulations. Next, since the fitted parameters are sufficient to describe the quasi-static behavior of the system, the theoretical elastic behavior of a single frustum is validated and investigated, utilizing finite element simulations. Finally, by estimating the remaining parameters, the overall dynamic behavior of the system is examined utilizing the experimental setup.

\subsection{Estimation of the model's parameters} \label{sec31}

The first stage of the parameter estimation process is finding the effective geometry of the channel system. For this, a set of 38 experiments where a ${V=90}$ ${\rm{cm}^3}$ vessel was filled through the channel system while the latter was detached from the straw, was executed. Assuming the flow is mostly steady, the flow rate is approximated as ${\rm{d}} V/{\rm{d}}t \approx V/t_f$, where $t_f$ denotes the time taken to fill the vessel. Under this assumption, the filling times $t_f$ and the known pressure differences \(\Delta p\left( t \right) \buildrel \Delta \over = {p_{{\rm{external}}}}\left( t \right) - p\left( t \right)\) spanning between 5 kPa and 90 kPa, are used to calculate the proportion coefficient that according to (\ref{eq2}) quadratically relates the volumetric flow rate to the pressure difference. Indeed, Figure \ref{FigureS1} in Appendix \ref{Appx1} shows that the coefficient achieved by curve fitting considering the theoretical Bernoulli model captures well the pressure-flow rate relation, thus it is utilized to compute the effective radius of the channel which is given by $r_{ch}$=0.52 mm while taking the fluid's density as $\rho=1000$ ${\rm{kg/m^3}}$.

As mentioned above, the experimental demonstrator includes a straw with nine multistable cells, meaning that under the assumption of a two-dimensional motion, the demonstrator should be described by 36 DOFs. However, since the computational effort needed to simulate a large number of equations is high, the DOFs of the odd frusta are disregarded meaning that they are considered rigid, which leaves the model with 18 DOFs. This assumption which is employed in the remaining steps of the parameter estimation process as well as the analyses presented in section \ref{sec33}, is supported by the observations that \({\zeta _{0,1}} \approx 1.66{\left| {{\zeta _{0,2}}} \right|}\) (see next paragraph), and that the thickness of the odd frusta is higher than this of the even ones, both significantly increase the stiffness of the odd frusta. This simplification is further justified, based on numerical analyses, as discussed in section \ref{sec32}.

The next stage is determining the geometrical parameters as well as the Young modulus of the solid. For this, first a caliper having a resolution of 0.01 mm was used to measure the inner and outer radii of the frusta, according to which $r_i=$11.1 mm and $r_o=$14.65 mm. Moreover, it was utilized to measure the axial lengths of 63 folded cells and 26 deployed cells, showing that the distance between the bases of a single folded cell is 1.5 mm whereas the corresponding value in a deployed cell is 5.8 mm. Further, the pressure values causing each active frustum to lose stability from its folded and deployed states while the straw is in a predetermined configuration, were assessed following a series of experiments. In these experiments, the pressure in which a frustum loses stability from a folded state was achieved by deploying the entire straw except for the frustum under investigation, followed by elevating the pressure in increments of 0.5 kPa until the frustum post-buckled to a different state. A similar procedure was executed to examine the pressure causing each frustum to lose stability from its deployed state. In this case all other frusta were folded, and the negative pressure value causing the frustum under investigation to post-buckle into a different state was evaluated. These experiments showed high consistency, where all active frusta except for the one farthest from the base lost stability from their deployed and folded states in approximately -13.5 kPa and 22.5 kPa, respectively. Thus, for simplicity the properties of all active frusta are considered identical. The main cause for the asymmetry between the pressure values leading to instability stems from the elastic properties of the frusta. This can be seen analytically from Eq. (\ref{eq6a}) and graphically from Figures \ref{Figure4} and \ref{Figure5} (both thoroughly discussed in section \ref{sec32}), which show that if the thickness of a frustum is not infinitely small its elastic behavior is not symmetrical around $\zeta_{n,k}=0$. Namely, $\zeta_{0,k}$ serves as a stable equilibrium state, whereas the absolute value of the second stable axisymmetric equilibrium is smaller than $\left| {{\zeta _{0,k}}} \right|$. Furthermore, Figure \ref{Figure5} clearly shows that among the axisymmetric stable equilibria, the one with the smaller deflection loses stability under a lower pressure variation. Consequently, since the absolute pressure value that leads to instability at the snap-up state is smaller than the corresponding value of the snap-down state, $\zeta_{0,2}$ must be negative. Therefore, according to the abovementioned axial lengths of the folded and deployed cells, ${\zeta _{0,1}} + {\zeta _{0,2}} = 1.5$ mm. Finally, to compute the missing parameters, the values specified above are used in a quasi-static analysis geared to find the pressure values that lead to the instabilities of the first active frustum. This analysis numerically describes the system according to the theoretical model while dropping all time-derivatives, yet it considers the effect of gravity, excluding the negligible torque originating in the offset between the sealing's center of mass, and the straw's closed end. Utilizing the nonlinear algebraic system achieved by these means, the stable states corresponding to the ones examined experimentally to assess the stability of the first active frustum, are computed for different pressures values. For each pressure, the eigenvalues of the $2 \times 2$ Jacobian which considers only the DOFs of the first active frustum and the corresponding equations are calculated to examine stability, where a non-positive eigenvalue indicates on instability. According to this analysis, the parameters leading to instability at the formerly mentioned pressure values and meet the geometrical constraints discussed above are: ${\zeta _{0,1}} = 3.78$ mm, ${\zeta _{0,2}} = -2.28$ mm, $h=0.508$ mm and $E=0.92$ GPa. Due to the model's simplifying assumptions, these should be referred to as effective values, rather than the correct physical values.

 The last two parameters to be calculated are the damping coefficients of the different DOFs, recalling that for simplicity the coefficients related to all axisymmetric DOFs are considered identical, as well as those related to the antisymmetric DOFs. The calibration of these parameters is thoroughly discussed in section \ref{sec33}, as part of the experimental verification of the theoretical model's overall dynamics.

\subsection{Numerical verification and investigation of the system’s elastic properties} \label{sec32}

\begin{figure}
\begin{center}
\includegraphics[width=\linewidth]{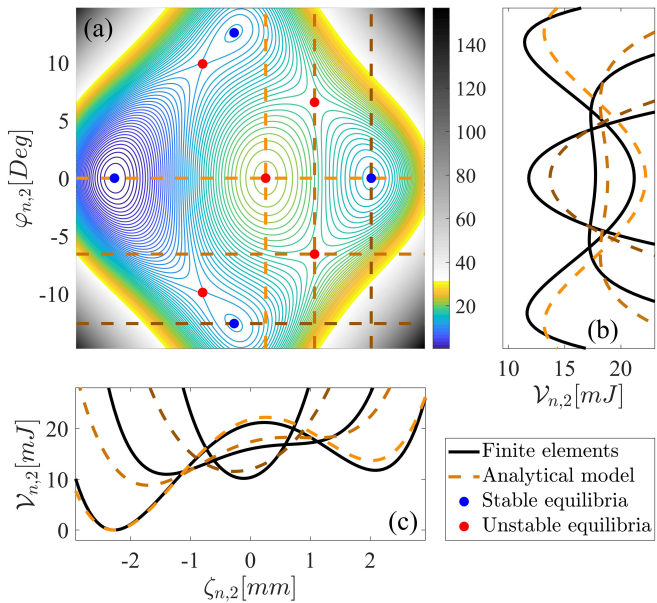}
\end{center}
\caption{(a) The theoretical potential energy of a single active frustum under zero gauge pressure according to (\ref{eq7}), alongside the stable and unstable equilibria. (b) and (c) Comparison between the theoretical values (orange dashed curves) and the numerically calculated values (black solid curves) of the potential energy, at the cutouts described by the orange dashed lines in panel (a).}
\label{Figure4}
\end{figure}

To start shedding light on the behavior of the system under investigation, this section begins with examining the elastic nature of a single active frustum while it is not subjected to pressure. This is done based on the theoretical formulation, and finite element analyses. Indeed, Figure \ref{Figure4} shows the strain energy function of a single active frustum based on Eq. (\ref{eq7}) utilizing the parameters specified above. This figure also presents the frustum's equilibrium states achieved by nulling Eq. (\ref{eq6a}) and Eq. (\ref{eq6b}), where their stability is classified based on the eigenvalues of the corresponding Jacobians. Namely, a certain equilibrium state is stable if both eigenvalues of the $2 \times 2$ partial derivative matrix of Eq. (\ref{eq6a}) and Eq. (\ref{eq6b}) with respect to the frustum's DOFs, are positive around this state. Figure \ref{Figure4} further shows several cutouts of the theoretical strain energy function where either the axisymmetric or the antisymmetric DOF is fixed. These are presented alongside the corresponding cutouts achieved from a finite element scheme devised in COMSOL Multiphyics, where the Poisson ratio which is absent from the theoretical model is taken arbitrarily as $\nu=0.33$. In this scheme, the frustum is discretized by second order rectangular shell elements modelled according to Reissner–Mindlin shell theory which considers shear deformations. These elements are distributed uniformly such that there are 15 elements in the radial direction and 400 elements along the circumference. The curves extracted from the finite element scheme are computed utilizing a built-in function applied on the results of numerous stationary simulations, in which the displacement of the frustum's small base along the $\hat z$ axis is dictated, whereas its large base is bounded to a plane in a way that allows free rotations in the tangential direction, in agreement with the model's boundary conditions. 

\begin{figure}
\begin{center}
\includegraphics[width=\linewidth]{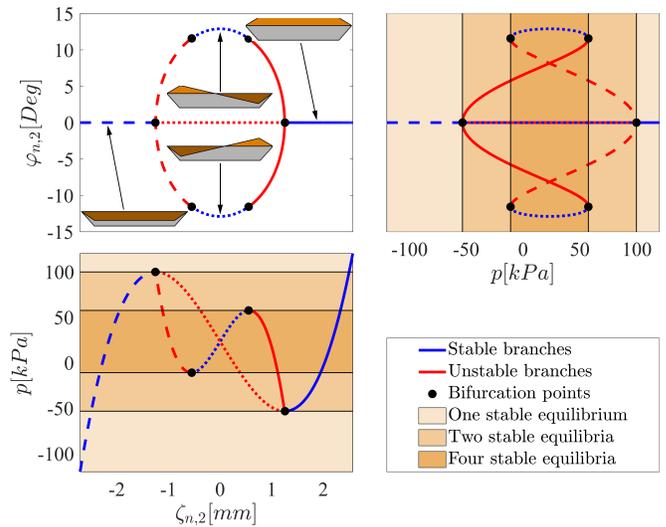}
\end{center}
\caption{Three projected views describing the dependency of the equilibrium states of a single elastic frustum, on the pressure applied on it directly. The red curves describe the unstable branches, whereas the blue curves represent the stable branches, corresponding to the states drawn on the $\zeta_{n,2}-\varphi_{n,2}$ plane.}
\label{Figure5}
\end{figure}

Figure \ref{Figure4} shows a very good agreement between the theoretical and the numerically obtained results in moderately small values of the DOFs. Yet, significant values of $\zeta_{n,2}$ and $\varphi_{n,2}$ lead to a weaker correlation as they result in significant base angles, which are assumed to be small in the model derivation. Nevertheless, the model manages to capture the multistability of the frustum, and its range of validity seems adequate to capture most reasonable deformations. Figure \ref{Figure4} further shows that when the frustum is not subjected to pressure it has nine equilibrium states, four of which are stable and correspond to folded and deployed axisymmetric states, as well as two antisymmetric states, see graphical demonstration in Figure \ref{Figure5}. These stable states provide a non-pressurized straw the capability of staying stable in a large number of equilibrium states where each active frustum can be in any of these four states. Finally, as formerly mentioned, Figure \ref{Figure4} shows that in the realistic case where the thickness of the frustum is not infinitely small, its potential energy and equilibrium states are not symmetrical around $\zeta_{n,k}=0$. 

To complete the picture of the quasi-static behavior of a single active frustum, Figure \ref{Figure5} shows its analytically obtained equilibrium states while it is subjected to different pressure values. This figure shows four bifurcation pressure values, separating five regions with different number of stable states. The first region corresponds to small absolute pressure values thus is described qualitatively by Figure \ref{Figure4}, meaning that in this region the frustum has four stable states. Conversely, in high positive gauge pressure values the frustum can stay stable only in a deployed state, and similarly in high negative pressure values the only stable state is the folded state. At the two supplementary regions achieved in moderate pressure values (positive or negative), the frustum has two stable equilibria, corresponding to both deployed and folded states. Finally, from Figure \ref{Figure4} and Figure \ref{Figure5} one should conclude that in slow motions where the inertial forces are weak, it is energetically preferable for the frustum to exhibit antisymmetrical motion, while switching between folded and deployed states. As seen in the upcoming section, in some cases it might pass through a stable antisymmetric equilibrium.

The last analysis executed which deals with the system's elastic properties examines the assumption that all odd frusta can be considered rigid. This analysis utilizes a finite element scheme established in COMSOL Multiphysics, which is similar to the one discussed above. Yet, in this case the analysis describes a complete multistable cell composed of two interconnected frusta with identical thicknesses, whose unstressed axisymmetric deflections are $\zeta_{0,1}$ and $\zeta_{0,2}$. In agreement with the model's assumptions, the bases of both frusta are allowed to rotate freely around the tangential direction where their large bases are forced to have identical translational motion. Further, the displacements between the small bases are dictated and denoted ${\zeta _{{\rm{dictated}}}}$ and ${\varphi _{{\rm{dictated}}}}$, where in similar to the scheme discussed above, each frustum is discretized by second order rectangular shell elements, dividing it to 15 radial and 400 tangential evenly distributed segments. Based on this scheme, Figure \ref{FigureS2} in Appendix \ref{Appx2} compares the idealized deformations of the cell represented only by those of the frustum with the smaller unstressed deflection, and the deformations of this frustum in the more realistic case where the second frustum is not considered rigid. In the latter case, the deflections of the relevant frustum are computed based on the relative deformations of its bases, utilizing line integration to describe only their rigid body motion. Figure \ref{FigureS2} shows that in moderate values of the DOFs which suit most practical deformations, the deviations between the realistic and the idealized cases are small, implying that the model simplification which truncates half of its DOFs, is indeed applicable.

\subsection{Experimental validation and investigation of the overall dynamics} \label{sec33}

To validate and examine the overall dynamic behavior of the theoretical model, the latter is simulated by a numerical scheme of its truncated form which consists of 18 coupled ordinary differential equations. These describe both axisymmetric and antisymmetric DOFs of all nine active frusta. In agreement with the experimental setup, each simulation starts from a stable state achieved by solving the static form of the governing equations which neglects all time dependencies, see discussion in section \ref{sec31}. Next, the dynamic part of the simulation is executed by numerical integration of the non-degenerated equations under a pressure excitation given by ${p_{{\rm{external}}}}\left( t \right) = {p_{ss}}\left[ {1 - \exp \left( { - 50\,\,{{\mathop{\rm s}\nolimits} ^{ - 1}} \cdot t} \right)} \right]$, bringing the input pressure to $97\%$ of the desired value denoted $p_{ss}$ after 70 $\rm{ms}$, in agreement with the rising time measured experimentally. The geometrical and physical parameters utilized in all numerical simulations are those specified above. However, recalling that the damping coefficients ${c_\zeta }$ and ${c_\varphi }$ which are affected by both structural and fluidic effects are yet to be quantified, these are manually fitted to best describe each experiment. Furthermore, to examine the assumption claiming that the fluidic effects related to the flow inside the water reservoir and the straw can be neglected, the effective radius of the channel is also readjusted, followed by comparison to its pre-calibrated value.

\begin{figure*}
\begin{center}
\includegraphics[width=1\linewidth]{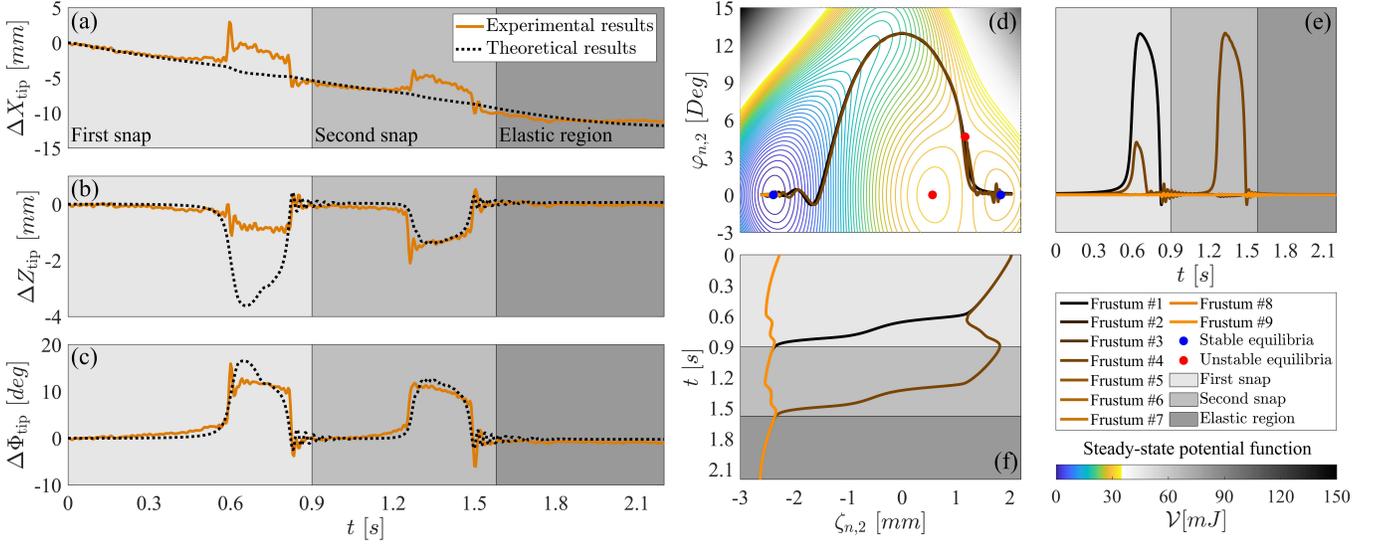}
\end{center}
\caption{(a,b,c) Comparison between the experimentally measured and theoretically simulated motion of the straw's free end along (a) the horizontal and (b) the vertical axes, as well as (c) its orientation, while being folded due to a steady-state external pressure of -25 kPa. (d) The theoretical dynamic responses of all frusta on the configuration space, and the potential energy of each frustum when subjected to the steady-state static pressure. (e,f) The theoretical time responses of (e) the axisymmetric and (f) the antisymmetric DOFs of all frusta.}
\label{Figure6}
\end{figure*}

\begin{figure*}
\begin{center}
\includegraphics[width=1\linewidth]{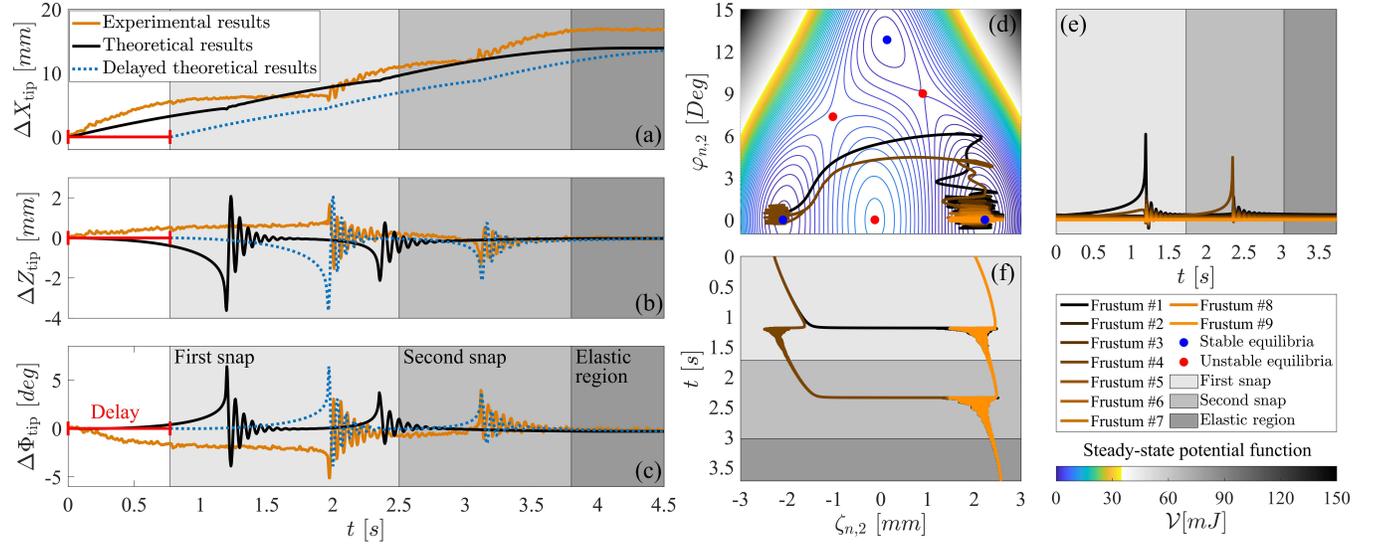}
\end{center}
\caption{(a,b,c) Comparison between the experimentally measured and theoretically simulated motion (as achieved originally and after a delay of 0.77 s) of the straw's free end along (a) the horizontal and (b) the vertical axes, as well as (c) its orientation, throughout deployment with an external steady-state pressure of 35 kPa. (d) The theoretical dynamic responses of all frusta on the configuration space, and the potential energy of each frustum when subjected to the steady-state static pressure. (e,f) The theoretical time responses of (e) the axisymmetric and (f) the antisymmetric DOFs of all frusta.}
\label{Figure7}
\end{figure*}

Figure \ref{Figure6} (a,b,c) compares the experimentally measured and theoretically simulated dynamic responses of the demonstrator's free end, while being folded by an external steady-state gauge pressure of $p_{ss}=-25$ kPa. Here, the experiment and corresponding simulation start from an initial state where the first and fourth multistable cells are deployed, and all others are folded. The manually fitted damping coefficients considered in the simulation are $c_{\zeta}=620$ $\rm{N \cdot s/m}$ and $c_{\varphi}=7 \times 10^{-4}$ $\rm{N \cdot m \cdot s/rad}$, whereas the effective radius of the channel is taken as $r_{ch}=0.43$ $\rm{mm}$, which is $82.5 \%$ of its pre-calibrated value. Similarly, Figure \ref{Figure7} (a,b,c) compares the measured and theoretical responses of the system's free end throughout deployment by an external steady-state gauge pressure of $p_{ss}=35$ kPa. Here, the system begins from an initial state where the first and fourth cells are folded, and all others are deployed. The damping coefficients utilized in this case are $c_{\zeta}=0.1$ $\rm{N \cdot s/m}$ and $c_{\varphi}=3 \times 10^{-3}$ $\rm{N \cdot m \cdot  s/rad}$, whereas the effective radius of the channel is taken as $r_{ch}=0.315$ $\rm{mm}$, which is $60.5 \%$ of its pre-calibrated value. Further, to analyse the internal motion of the system, Figure \ref{Figure6} (d,e,f) and Figure \ref{Figure7} (d,e,f) show the numerically simulated theoretical responses of the different DOFs in time, and on the configuration space alongside the potential functions of the different frusta considering the steady-state pressure. Finally, the video in the supporting information, whose caption appears in Appendix \ref{Appx4}, shows the theoretically simulated and the experimentally obtained responses, corresponding to Figure \ref{Figure6} and Figure \ref{Figure7}.

Figure \ref{Figure6} shows that the model well describes the dynamic behavior of the system which folds throughout two steps, where in each step one of the deployed frusta snaps through an antisymmetric stable state, into a fully folded state, see panels (d,e,f). This figure further shows an overall high quantitative correlation between the experiment and the theory, as the model manages to capture both the magnitude and time scales of the rotational and horizontal motion of the straw's free end. However, the theoretical motion along the vertical axis is in a lesser agreement with the corresponding experimental response, which is reflected in a strong overestimation of the magnitude throughout the first snap-through post-buckling. One reason for this overestimation is that even though the torque applied to the first active frustum is higher than this applied to the fourth active frustum, in practice the first one to post-buckle is the latter due to non-uniformity in the physical properties of the different frusta. Other possible explanations of the deviations are unmodeled effects such as longitudinal curvatures of the unloaded straw, originating from internal stresses. Finally, according to Figure \ref{Figure6} (d) showing the theoretical dynamic responses of the active frusta on the configuration space, the folding of the initially deployed frusta hardly deviates from the upper bifurcation branch presented in Figure \ref{Figure5}. This indicates that the behavior of the system is governed by its potential energy, which implies on weak inertial effects.

Next, Figure \ref{Figure7} shows that the significantly lower damping coefficient related to the axisymmetric DOFs, achieved when deploying the system leads to a radically different dynamic behavior. In this case the system's inertial forces are much more significant compared to those achieved while the system is being folded. Consequently, each frustum snaps from a folded state to a deployed state with only a slight antisymmetric motion and without passing through an intermediate state. Figure \ref{Figure7} further shows that the model accurately describes the magnitude of both translational and rotational motions of the straw's closed end, as well as the frequency and decay rates of the oscillations occurring after each snap-through post-buckling. However, even though the model also manages to capture the time-period between both snaps, it significantly underestimates the time until the first one. Nevertheless, several repetitions of the same experiment show almost identical dynamic responses, where in each repetition the initial time period is shortened, see Figure \ref{FigureS3} in Appendix \ref{Appx3}. The latter implies that this deviation originates from unmodeled effects such as cohesion forces acting between the folded frusta, or non-negligible stiffness of the thin creases connecting them, which weaken in each cycle. As seen in Figure \ref{Figure7}, a delay of 0.77 s in the theoretical dynamic response compensates on the initial period underestimation, thus leads to a good correlation with the measurements. Yet, it can be seen that despite the effect of gravity, throughout the first snap, the frustum that underwent post-buckling bent upwards since it is not perfectly axisymmetric. Further, as already mentioned, since the fourth active frustum is weaker than the first one, in contrast to the theoretically predicted response, it was the first one to undergo post-buckling.

Recalling that the theoretical responses presented in Figure \ref{Figure6} and Figure \ref{Figure7} are achieved utilizing different damping coefficients and equivalent radii of the channel, the model can be adjusted to different configurations and responses merely by fitting the parameters that are affected by fluidic effects. Thus, to reduce the number of parameters to be calibrated, theoretical modeling of the fluid in both the straw and the water reservoir should be added to the model derived in section \ref{sec2}. A possible fluidic effect can be thin film damping related to the fluid trapped between closely adjacent frusta in partially- or fully-folded multistable cells. Other possible effects can originate from inertial forces of the fluid in both the water reservoir and the straw, which increase the resistance to folding and deployment of the latter, and consequently reduce the channel's effective radius.

\section{Conclusion}
The large number of complex stable equilibrium states attainable by straw-like elements make them promising reconfigurable structures. Thus, the ability to predict their dynamics has a great potential for countless practical applications, spanning from deployable space structures to soft robots. Indeed, we derived a comprehensive model describing the dynamics of fluid-filled straw-like elements, consisting of interconnected multistable building blocks, each of which has up to four stable equilibria. After the theoretical derivation, we followed an algorithm geared to efficiently estimate most parameters of the system, for sake of adjusting the model to a specific experimental setup. Next, we validated the theoretical quasi-static behavior of the constituent building blocks compared to finite element analyses, and numerically studied this behavior to gain physical insights regarding the system's dynamics when inertial effects are weak. Finally, we experimentally showed that by calibrating a few parameters which are influenced by fluidic effects, it is possible to capture various dynamic behaviors achieved under different conditions. The latter implies that the theoretical model captures well all the dominant effects of the highly complex system studied here, except for those related to the fluid which we modelled in a simplistic manner. Nevertheless, either a simple calibration, or a rigorous supplementary derivation of these effects can adjust the model to different configurations that exhibit distinct fluidic phenomena, making the suggested model universal. The theoretical model can be further explored to gain a deeper understanding of the dynamics of straw-like elements, and it can be leveraged to describe structures with higher complexity, consisting of hierarchical configurations of such elements.

\appendix

\section{Identification of the channel's effective radius} \label{Appx1}
\begin{figure}
\begin{center}
\includegraphics[width=\linewidth]{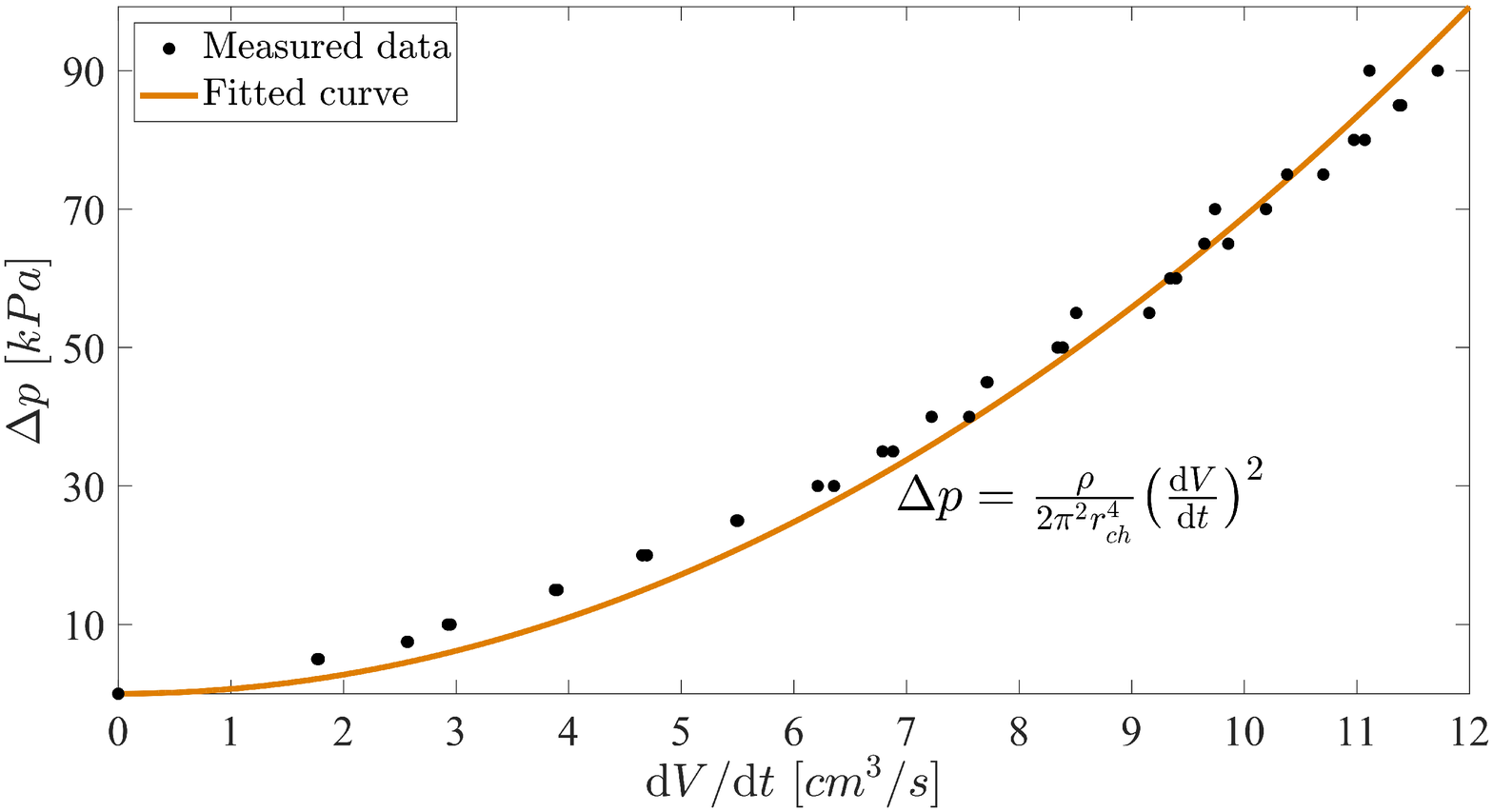}
\end{center}
\caption{The measured relation between the dictated pressure difference among the channel system's ends, and the time-averaged flow rate in absence of the sealed straw, alongside the fitted Bernoulli model.}
\label{FigureS1}
\end{figure}

Figure \ref{FigureS1} which is discussed in section \ref{sec31}, shows the experimentally obtained relation between the time-averaged flow rate in the channel system, and the dictated pressure difference between its ends, while the straw was detached from the system. The raw data presented in this figure is utilized to calculate the effective radius of the channel by the shown fitted steady Bernoulli formulation.

\section{The validity of the rigidization of the odd frusta} \label{Appx2}
\begin{figure}
\begin{center}
\includegraphics[width=\linewidth]{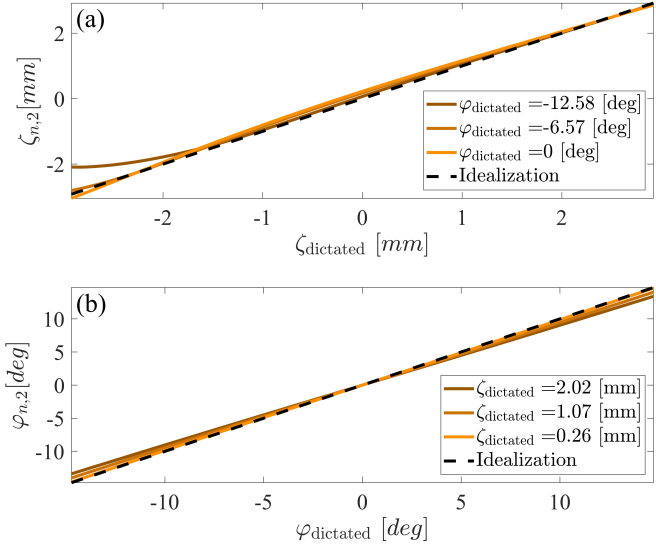}
\end{center}
\caption{The deformations of a single active frustum under dictated relative displacements between the bases of a whole multistable cell, according to finite-element analyses, and based on the idealization under which the frustum with the larger unstressed axisymmetric deflection can be considered rigid. The presented comparison is along (a) the horizontal and (b) the vertical cutouts of Figure \ref{Figure4}.}
\label{FigureS2}
\end{figure}

Figure \ref{FigureS2} mentioned in section \ref{sec32}, displays the deformations of the frustum with the smaller unstressed axisymmetric deflection, under dictated relative displacements between the bases of a whole multistable cell. The results are based on finite element analyses, and the simplification which claims that the deformations of the frustum that has a larger unstressed axisymmetric deflection, can be disregarded.

\section{Comparison between similar dynamic experiments} \label{Appx3}
\begin{figure}
\begin{center}
\includegraphics[width=\linewidth]{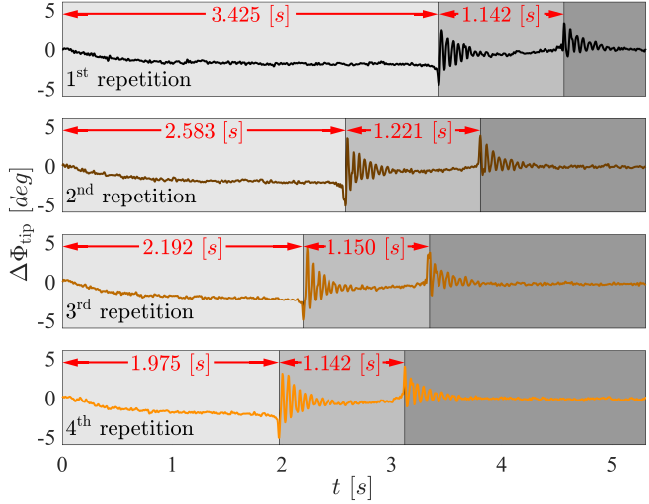}
\end{center}
\caption{The orientation of the straw's free end, the period between the pressure elevation and the first snap, and the time between the two snaps, in four repetitions of the experiment in which the first and fourth active frusta are deployed by a steady-state external input pressure of 35 kPa.}
\label{FigureS3}
\end{figure}

Figure \ref{FigureS3} shows four repetitions of the same dynamic deployment experiment, used to examine the claim that the period between the first and second snaps hardly changed, see section \ref{sec33}.

\section{Supplementary video} \label{Appx4}
The video that appears as supplementary information, shows the comparison between the theoretically predicted and the experimentally obtained dynamics of the system throughout folding and deploying of two distinct frusta. These responses correspond to Figure \ref{Figure6} and Figure \ref{Figure7}, presented and discussed in \ref{sec33}

\bibliography{References}

\begin{thebibliography}{24}%
\makeatletter
\providecommand \@ifxundefined [1]{%
 \@ifx{#1\undefined}
}%
\providecommand \@ifnum [1]{%
 \ifnum #1\expandafter \@firstoftwo
 \else \expandafter \@secondoftwo
 \fi
}%
\providecommand \@ifx [1]{%
 \ifx #1\expandafter \@firstoftwo
 \else \expandafter \@secondoftwo
 \fi
}%
\providecommand \natexlab [1]{#1}%
\providecommand \enquote  [1]{``#1''}%
\providecommand \bibnamefont  [1]{#1}%
\providecommand \bibfnamefont [1]{#1}%
\providecommand \citenamefont [1]{#1}%
\providecommand \href@noop [0]{\@secondoftwo}%
\providecommand \href [0]{\begingroup \@sanitize@url \@href}%
\providecommand \@href[1]{\@@startlink{#1}\@@href}%
\providecommand \@@href[1]{\endgroup#1\@@endlink}%
\providecommand \@sanitize@url [0]{\catcode `\\12\catcode `\$12\catcode
  `\&12\catcode `\#12\catcode `\^12\catcode `\_12\catcode `\%12\relax}%
\providecommand \@@startlink[1]{}%
\providecommand \@@endlink[0]{}%
\providecommand \url  [0]{\begingroup\@sanitize@url \@url }%
\providecommand \@url [1]{\endgroup\@href {#1}{\urlprefix }}%
\providecommand \urlprefix  [0]{URL }%
\providecommand \Eprint [0]{\href }%
\providecommand \doibase [0]{https://doi.org/}%
\providecommand \selectlanguage [0]{\@gobble}%
\providecommand \bibinfo  [0]{\@secondoftwo}%
\providecommand \bibfield  [0]{\@secondoftwo}%
\providecommand \translation [1]{[#1]}%
\providecommand \BibitemOpen [0]{}%
\providecommand \bibitemStop [0]{}%
\providecommand \bibitemNoStop [0]{.\EOS\space}%
\providecommand \EOS [0]{\spacefactor3000\relax}%
\providecommand \BibitemShut  [1]{\csname bibitem#1\endcsname}%
\let\auto@bib@innerbib\@empty
\bibitem [{\citenamefont {Bertoldi}\ \emph {et~al.}(2017)\citenamefont
  {Bertoldi}, \citenamefont {Vitelli}, \citenamefont {Christensen},\ and\
  \citenamefont {Van~Hecke}}]{bertoldi2017flexible}%
  \BibitemOpen
  \bibfield  {author} {\bibinfo {author} {\bibfnamefont {K.}~\bibnamefont
  {Bertoldi}}, \bibinfo {author} {\bibfnamefont {V.}~\bibnamefont {Vitelli}},
  \bibinfo {author} {\bibfnamefont {J.}~\bibnamefont {Christensen}},\ and\
  \bibinfo {author} {\bibfnamefont {M.}~\bibnamefont {Van~Hecke}},\ }\bibfield
  {title} {\bibinfo {title} {Flexible mechanical metamaterials},\ }\href@noop
  {} {\bibfield  {journal} {\bibinfo  {journal} {Nature Reviews Materials}\
  }\textbf {\bibinfo {volume} {2}},\ \bibinfo {pages} {1} (\bibinfo {year}
  {2017})}\BibitemShut {NoStop}%
\bibitem [{\citenamefont {Yang}\ and\ \citenamefont
  {Ma}(2019)}]{yang2019multi}%
  \BibitemOpen
  \bibfield  {author} {\bibinfo {author} {\bibfnamefont {H.}~\bibnamefont
  {Yang}}\ and\ \bibinfo {author} {\bibfnamefont {L.}~\bibnamefont {Ma}},\
  }\bibfield  {title} {\bibinfo {title} {Multi-stable mechanical metamaterials
  by elastic buckling instability},\ }\href@noop {} {\bibfield  {journal}
  {\bibinfo  {journal} {Journal of materials science}\ }\textbf {\bibinfo
  {volume} {54}},\ \bibinfo {pages} {3509} (\bibinfo {year}
  {2019})}\BibitemShut {NoStop}%
\bibitem [{\citenamefont {Hua}\ \emph {et~al.}(2019)\citenamefont {Hua},
  \citenamefont {Lei}, \citenamefont {Zhang}, \citenamefont {Gao},\ and\
  \citenamefont {Fang}}]{hua2019multistable}%
  \BibitemOpen
  \bibfield  {author} {\bibinfo {author} {\bibfnamefont {J.}~\bibnamefont
  {Hua}}, \bibinfo {author} {\bibfnamefont {H.}~\bibnamefont {Lei}}, \bibinfo
  {author} {\bibfnamefont {Z.}~\bibnamefont {Zhang}}, \bibinfo {author}
  {\bibfnamefont {C.}~\bibnamefont {Gao}},\ and\ \bibinfo {author}
  {\bibfnamefont {D.}~\bibnamefont {Fang}},\ }\bibfield  {title} {\bibinfo
  {title} {Multistable cylindrical mechanical metastructures: theoretical and
  experimental studies},\ }\href@noop {} {\bibfield  {journal} {\bibinfo
  {journal} {Journal of Applied Mechanics}\ }\textbf {\bibinfo {volume} {86}}
  (\bibinfo {year} {2019})}\BibitemShut {NoStop}%
\bibitem [{\citenamefont {Findeisen}\ \emph {et~al.}(2017)\citenamefont
  {Findeisen}, \citenamefont {Hohe}, \citenamefont {Kadic},\ and\ \citenamefont
  {Gumbsch}}]{findeisen2017characteristics}%
  \BibitemOpen
  \bibfield  {author} {\bibinfo {author} {\bibfnamefont {C.}~\bibnamefont
  {Findeisen}}, \bibinfo {author} {\bibfnamefont {J.}~\bibnamefont {Hohe}},
  \bibinfo {author} {\bibfnamefont {M.}~\bibnamefont {Kadic}},\ and\ \bibinfo
  {author} {\bibfnamefont {P.}~\bibnamefont {Gumbsch}},\ }\bibfield  {title}
  {\bibinfo {title} {Characteristics of mechanical metamaterials based on
  buckling elements},\ }\href@noop {} {\bibfield  {journal} {\bibinfo
  {journal} {Journal of the Mechanics and Physics of Solids}\ }\textbf
  {\bibinfo {volume} {102}},\ \bibinfo {pages} {151} (\bibinfo {year}
  {2017})}\BibitemShut {NoStop}%
\bibitem [{\citenamefont {Zhang}\ \emph {et~al.}(2020)\citenamefont {Zhang},
  \citenamefont {Wang}, \citenamefont {Tichem},\ and\ \citenamefont {van
  Keulen}}]{zhang2020design}%
  \BibitemOpen
  \bibfield  {author} {\bibinfo {author} {\bibfnamefont {Y.}~\bibnamefont
  {Zhang}}, \bibinfo {author} {\bibfnamefont {Q.}~\bibnamefont {Wang}},
  \bibinfo {author} {\bibfnamefont {M.}~\bibnamefont {Tichem}},\ and\ \bibinfo
  {author} {\bibfnamefont {F.}~\bibnamefont {van Keulen}},\ }\bibfield  {title}
  {\bibinfo {title} {Design and characterization of multi-stable mechanical
  metastructures with level and tilted stable configurations},\ }\href@noop {}
  {\bibfield  {journal} {\bibinfo  {journal} {Extreme Mechanics Letters}\
  }\textbf {\bibinfo {volume} {34}},\ \bibinfo {pages} {100593} (\bibinfo
  {year} {2020})}\BibitemShut {NoStop}%
\bibitem [{\citenamefont {Meaud}\ and\ \citenamefont
  {Che}(2017)}]{meaud2017tuning}%
  \BibitemOpen
  \bibfield  {author} {\bibinfo {author} {\bibfnamefont {J.}~\bibnamefont
  {Meaud}}\ and\ \bibinfo {author} {\bibfnamefont {K.}~\bibnamefont {Che}},\
  }\bibfield  {title} {\bibinfo {title} {Tuning elastic wave propagation in
  multistable architected materials},\ }\href@noop {} {\bibfield  {journal}
  {\bibinfo  {journal} {International Journal of Solids and Structures}\
  }\textbf {\bibinfo {volume} {122}},\ \bibinfo {pages} {69} (\bibinfo {year}
  {2017})}\BibitemShut {NoStop}%
\bibitem [{\citenamefont {Raney}\ \emph {et~al.}(2016)\citenamefont {Raney},
  \citenamefont {Nadkarni}, \citenamefont {Daraio}, \citenamefont {Kochmann},
  \citenamefont {Lewis},\ and\ \citenamefont {Bertoldi}}]{raney2016stable}%
  \BibitemOpen
  \bibfield  {author} {\bibinfo {author} {\bibfnamefont {J.~R.}\ \bibnamefont
  {Raney}}, \bibinfo {author} {\bibfnamefont {N.}~\bibnamefont {Nadkarni}},
  \bibinfo {author} {\bibfnamefont {C.}~\bibnamefont {Daraio}}, \bibinfo
  {author} {\bibfnamefont {D.~M.}\ \bibnamefont {Kochmann}}, \bibinfo {author}
  {\bibfnamefont {J.~A.}\ \bibnamefont {Lewis}},\ and\ \bibinfo {author}
  {\bibfnamefont {K.}~\bibnamefont {Bertoldi}},\ }\bibfield  {title} {\bibinfo
  {title} {Stable propagation of mechanical signals in soft media using stored
  elastic energy},\ }\href@noop {} {\bibfield  {journal} {\bibinfo  {journal}
  {Proceedings of the National Academy of Sciences}\ }\textbf {\bibinfo
  {volume} {113}},\ \bibinfo {pages} {9722} (\bibinfo {year}
  {2016})}\BibitemShut {NoStop}%
\bibitem [{\citenamefont {Nadkarni}\ \emph {et~al.}(2016)\citenamefont
  {Nadkarni}, \citenamefont {Arrieta}, \citenamefont {Chong}, \citenamefont
  {Kochmann},\ and\ \citenamefont {Daraio}}]{nadkarni2016unidirectional}%
  \BibitemOpen
  \bibfield  {author} {\bibinfo {author} {\bibfnamefont {N.}~\bibnamefont
  {Nadkarni}}, \bibinfo {author} {\bibfnamefont {A.~F.}\ \bibnamefont
  {Arrieta}}, \bibinfo {author} {\bibfnamefont {C.}~\bibnamefont {Chong}},
  \bibinfo {author} {\bibfnamefont {D.~M.}\ \bibnamefont {Kochmann}},\ and\
  \bibinfo {author} {\bibfnamefont {C.}~\bibnamefont {Daraio}},\ }\bibfield
  {title} {\bibinfo {title} {Unidirectional transition waves in bistable
  lattices},\ }\href@noop {} {\bibfield  {journal} {\bibinfo  {journal}
  {Physical review letters}\ }\textbf {\bibinfo {volume} {116}},\ \bibinfo
  {pages} {244501} (\bibinfo {year} {2016})}\BibitemShut {NoStop}%
\bibitem [{\citenamefont {Khajehtourian}\ and\ \citenamefont
  {Kochmann}(2020)}]{khajehtourian2020phase}%
  \BibitemOpen
  \bibfield  {author} {\bibinfo {author} {\bibfnamefont {R.}~\bibnamefont
  {Khajehtourian}}\ and\ \bibinfo {author} {\bibfnamefont {D.~M.}\ \bibnamefont
  {Kochmann}},\ }\bibfield  {title} {\bibinfo {title} {Phase transformations in
  substrate-free dissipative multistable metamaterials},\ }\href@noop {}
  {\bibfield  {journal} {\bibinfo  {journal} {Extreme Mechanics Letters}\ ,\
  \bibinfo {pages} {100700}} (\bibinfo {year} {2020})}\BibitemShut {NoStop}%
\bibitem [{\citenamefont {Jin}\ \emph {et~al.}(2020)\citenamefont {Jin},
  \citenamefont {Khajehtourian}, \citenamefont {Mueller}, \citenamefont
  {Rafsanjani}, \citenamefont {Tournat}, \citenamefont {Bertoldi},\ and\
  \citenamefont {Kochmann}}]{jin2020guided}%
  \BibitemOpen
  \bibfield  {author} {\bibinfo {author} {\bibfnamefont {L.}~\bibnamefont
  {Jin}}, \bibinfo {author} {\bibfnamefont {R.}~\bibnamefont {Khajehtourian}},
  \bibinfo {author} {\bibfnamefont {J.}~\bibnamefont {Mueller}}, \bibinfo
  {author} {\bibfnamefont {A.}~\bibnamefont {Rafsanjani}}, \bibinfo {author}
  {\bibfnamefont {V.}~\bibnamefont {Tournat}}, \bibinfo {author} {\bibfnamefont
  {K.}~\bibnamefont {Bertoldi}},\ and\ \bibinfo {author} {\bibfnamefont
  {D.~M.}\ \bibnamefont {Kochmann}},\ }\bibfield  {title} {\bibinfo {title}
  {Guided transition waves in multistable mechanical metamaterials},\
  }\href@noop {} {\bibfield  {journal} {\bibinfo  {journal} {Proceedings of the
  National Academy of Sciences}\ }\textbf {\bibinfo {volume} {117}},\ \bibinfo
  {pages} {2319} (\bibinfo {year} {2020})}\BibitemShut {NoStop}%
\bibitem [{\citenamefont {Katz}\ and\ \citenamefont
  {Givli}(2018)}]{katz2018solitary}%
  \BibitemOpen
  \bibfield  {author} {\bibinfo {author} {\bibfnamefont {S.}~\bibnamefont
  {Katz}}\ and\ \bibinfo {author} {\bibfnamefont {S.}~\bibnamefont {Givli}},\
  }\bibfield  {title} {\bibinfo {title} {Solitary waves in a bistable
  lattice},\ }\href@noop {} {\bibfield  {journal} {\bibinfo  {journal} {Extreme
  Mechanics Letters}\ }\textbf {\bibinfo {volume} {22}},\ \bibinfo {pages}
  {106} (\bibinfo {year} {2018})}\BibitemShut {NoStop}%
\bibitem [{\citenamefont {Ben-Haim}\ \emph {et~al.}(2020)\citenamefont
  {Ben-Haim}, \citenamefont {Salem}, \citenamefont {Or},\ and\ \citenamefont
  {Gat}}]{ben2020single}%
  \BibitemOpen
  \bibfield  {author} {\bibinfo {author} {\bibfnamefont {E.}~\bibnamefont
  {Ben-Haim}}, \bibinfo {author} {\bibfnamefont {L.}~\bibnamefont {Salem}},
  \bibinfo {author} {\bibfnamefont {Y.}~\bibnamefont {Or}},\ and\ \bibinfo
  {author} {\bibfnamefont {A.~D.}\ \bibnamefont {Gat}},\ }\bibfield  {title}
  {\bibinfo {title} {Single-input control of multiple fluid-driven elastic
  actuators via interaction between bistability and viscosity},\ }\href@noop {}
  {\bibfield  {journal} {\bibinfo  {journal} {Soft Robotics}\ }\textbf
  {\bibinfo {volume} {7}},\ \bibinfo {pages} {259} (\bibinfo {year}
  {2020})}\BibitemShut {NoStop}%
\bibitem [{\citenamefont {Glozman}\ \emph {et~al.}(2010)\citenamefont
  {Glozman}, \citenamefont {Hassidov}, \citenamefont {Senesh},\ and\
  \citenamefont {Shoham}}]{glozman2010self}%
  \BibitemOpen
  \bibfield  {author} {\bibinfo {author} {\bibfnamefont {D.}~\bibnamefont
  {Glozman}}, \bibinfo {author} {\bibfnamefont {N.}~\bibnamefont {Hassidov}},
  \bibinfo {author} {\bibfnamefont {M.}~\bibnamefont {Senesh}},\ and\ \bibinfo
  {author} {\bibfnamefont {M.}~\bibnamefont {Shoham}},\ }\bibfield  {title}
  {\bibinfo {title} {A self-propelled inflatable earthworm-like endoscope
  actuated by single supply line},\ }\href@noop {} {\bibfield  {journal}
  {\bibinfo  {journal} {IEEE Transactions on Biomedical Engineering}\ }\textbf
  {\bibinfo {volume} {57}},\ \bibinfo {pages} {1264} (\bibinfo {year}
  {2010})}\BibitemShut {NoStop}%
\bibitem [{\citenamefont {Peretz}\ \emph {et~al.}(2020)\citenamefont {Peretz},
  \citenamefont {Mishra}, \citenamefont {Shepherd},\ and\ \citenamefont
  {Gat}}]{peretz2020underactuated}%
  \BibitemOpen
  \bibfield  {author} {\bibinfo {author} {\bibfnamefont {O.}~\bibnamefont
  {Peretz}}, \bibinfo {author} {\bibfnamefont {A.~K.}\ \bibnamefont {Mishra}},
  \bibinfo {author} {\bibfnamefont {R.~F.}\ \bibnamefont {Shepherd}},\ and\
  \bibinfo {author} {\bibfnamefont {A.~D.}\ \bibnamefont {Gat}},\ }\bibfield
  {title} {\bibinfo {title} {Underactuated fluidic control of a continuous
  multistable membrane},\ }\href@noop {} {\bibfield  {journal} {\bibinfo
  {journal} {Proceedings of the National Academy of Sciences}\ }\textbf
  {\bibinfo {volume} {117}},\ \bibinfo {pages} {5217} (\bibinfo {year}
  {2020})}\BibitemShut {NoStop}%
\bibitem [{\citenamefont {Bende}\ \emph {et~al.}(2018)\citenamefont {Bende},
  \citenamefont {Yu}, \citenamefont {Corbin}, \citenamefont {Dias},
  \citenamefont {Santangelo}, \citenamefont {Hanna},\ and\ \citenamefont
  {Hayward}}]{bende2018overcurvature}%
  \BibitemOpen
  \bibfield  {author} {\bibinfo {author} {\bibfnamefont {N.~P.}\ \bibnamefont
  {Bende}}, \bibinfo {author} {\bibfnamefont {T.}~\bibnamefont {Yu}}, \bibinfo
  {author} {\bibfnamefont {N.~A.}\ \bibnamefont {Corbin}}, \bibinfo {author}
  {\bibfnamefont {M.~A.}\ \bibnamefont {Dias}}, \bibinfo {author}
  {\bibfnamefont {C.~D.}\ \bibnamefont {Santangelo}}, \bibinfo {author}
  {\bibfnamefont {J.~A.}\ \bibnamefont {Hanna}},\ and\ \bibinfo {author}
  {\bibfnamefont {R.~C.}\ \bibnamefont {Hayward}},\ }\bibfield  {title}
  {\bibinfo {title} {Overcurvature induced multistability of linked conical
  frusta: how a ‘bendy straw’holds its shape},\ }\href@noop {} {\bibfield
  {journal} {\bibinfo  {journal} {Soft matter}\ }\textbf {\bibinfo {volume}
  {14}},\ \bibinfo {pages} {8636} (\bibinfo {year} {2018})}\BibitemShut
  {NoStop}%
\bibitem [{\citenamefont {Bernardes}\ and\ \citenamefont
  {Viollet}(2022)}]{bernardes2022design}%
  \BibitemOpen
  \bibfield  {author} {\bibinfo {author} {\bibfnamefont {E.}~\bibnamefont
  {Bernardes}}\ and\ \bibinfo {author} {\bibfnamefont {S.}~\bibnamefont
  {Viollet}},\ }\bibfield  {title} {\bibinfo {title} {Design of an origami
  bendy straw for robotic multistable structures},\ }\href@noop {} {\bibfield
  {journal} {\bibinfo  {journal} {Journal of Mechanical Design}\ }\textbf
  {\bibinfo {volume} {144}} (\bibinfo {year} {2022})}\BibitemShut {NoStop}%
\bibitem [{\citenamefont {Pan}\ \emph {et~al.}(2019)\citenamefont {Pan},
  \citenamefont {Li}, \citenamefont {Li}, \citenamefont {Yang}, \citenamefont
  {Liu},\ and\ \citenamefont {Chen}}]{pan20193d}%
  \BibitemOpen
  \bibfield  {author} {\bibinfo {author} {\bibfnamefont {F.}~\bibnamefont
  {Pan}}, \bibinfo {author} {\bibfnamefont {Y.}~\bibnamefont {Li}}, \bibinfo
  {author} {\bibfnamefont {Z.}~\bibnamefont {Li}}, \bibinfo {author}
  {\bibfnamefont {J.}~\bibnamefont {Yang}}, \bibinfo {author} {\bibfnamefont
  {B.}~\bibnamefont {Liu}},\ and\ \bibinfo {author} {\bibfnamefont
  {Y.}~\bibnamefont {Chen}},\ }\bibfield  {title} {\bibinfo {title} {3d pixel
  mechanical metamaterials},\ }\href@noop {} {\bibfield  {journal} {\bibinfo
  {journal} {Advanced Materials}\ }\textbf {\bibinfo {volume} {31}},\ \bibinfo
  {pages} {1900548} (\bibinfo {year} {2019})}\BibitemShut {NoStop}%
\bibitem [{\citenamefont {Almen}\ and\ \citenamefont
  {Laszlo}(1936)}]{almen1936uniform}%
  \BibitemOpen
  \bibfield  {author} {\bibinfo {author} {\bibfnamefont {J.~O.}\ \bibnamefont
  {Almen}}\ and\ \bibinfo {author} {\bibfnamefont {A.}~\bibnamefont {Laszlo}},\
  }\bibfield  {title} {\bibinfo {title} {The uniform-section disk spring},\
  }\href@noop {} {\bibfield  {journal} {\bibinfo  {journal} {Transactions of
  the American Society of Mechanical Engineers}\ }\textbf {\bibinfo {volume}
  {58}},\ \bibinfo {pages} {305} (\bibinfo {year} {1936})}\BibitemShut
  {NoStop}%
\bibitem [{\citenamefont {Ilssar}\ and\ \citenamefont
  {Gat}(2020)}]{ilssar2020inflation}%
  \BibitemOpen
  \bibfield  {author} {\bibinfo {author} {\bibfnamefont {D.}~\bibnamefont
  {Ilssar}}\ and\ \bibinfo {author} {\bibfnamefont {A.~D.}\ \bibnamefont
  {Gat}},\ }\bibfield  {title} {\bibinfo {title} {On the inflation and
  deflation dynamics of liquid-filled, hyperelastic balloons},\ }\href@noop {}
  {\bibfield  {journal} {\bibinfo  {journal} {Journal of Fluids and
  Structures}\ }\textbf {\bibinfo {volume} {94}},\ \bibinfo {pages} {102936}
  (\bibinfo {year} {2020})}\BibitemShut {NoStop}%
\bibitem [{\citenamefont {Mastricola}\ \emph {et~al.}(2017)\citenamefont
  {Mastricola}, \citenamefont {Dreyer},\ and\ \citenamefont
  {Singh}}]{mastricola2017analytical}%
  \BibitemOpen
  \bibfield  {author} {\bibinfo {author} {\bibfnamefont {N.~P.}\ \bibnamefont
  {Mastricola}}, \bibinfo {author} {\bibfnamefont {J.~T.}\ \bibnamefont
  {Dreyer}},\ and\ \bibinfo {author} {\bibfnamefont {R.}~\bibnamefont
  {Singh}},\ }\bibfield  {title} {\bibinfo {title} {Analytical and experimental
  characterization of nonlinear coned disk springs with focus on edge friction
  contribution to force-deflection hysteresis},\ }\href@noop {} {\bibfield
  {journal} {\bibinfo  {journal} {Mechanical Systems and Signal Processing}\
  }\textbf {\bibinfo {volume} {91}},\ \bibinfo {pages} {215} (\bibinfo {year}
  {2017})}\BibitemShut {NoStop}%
\bibitem [{\citenamefont {La~Rosa}\ \emph {et~al.}(2001)\citenamefont
  {La~Rosa}, \citenamefont {Messina},\ and\ \citenamefont
  {Risitano}}]{la2001stiffness}%
  \BibitemOpen
  \bibfield  {author} {\bibinfo {author} {\bibfnamefont {G.}~\bibnamefont
  {La~Rosa}}, \bibinfo {author} {\bibfnamefont {M.}~\bibnamefont {Messina}},\
  and\ \bibinfo {author} {\bibfnamefont {A.}~\bibnamefont {Risitano}},\
  }\bibfield  {title} {\bibinfo {title} {Stiffness of variable thickness
  belleville springs},\ }\href@noop {} {\bibfield  {journal} {\bibinfo
  {journal} {J. Mech. Des.}\ }\textbf {\bibinfo {volume} {123}},\ \bibinfo
  {pages} {294} (\bibinfo {year} {2001})}\BibitemShut {NoStop}%
\bibitem [{\citenamefont {Timoshenko}(1956)}]{timoshenko1956strength}%
  \BibitemOpen
  \bibfield  {author} {\bibinfo {author} {\bibfnamefont {S.}~\bibnamefont
  {Timoshenko}},\ }\href@noop {} {\emph {\bibinfo {title} {Strength of
  Materials: Part II, Advanced Theory and Problems}}}\ (\bibinfo  {publisher}
  {van Nostrand},\ \bibinfo {year} {1956})\BibitemShut {NoStop}%
\bibitem [{\citenamefont {Reddy}(2017)}]{reddy2017energy}%
  \BibitemOpen
  \bibfield  {author} {\bibinfo {author} {\bibfnamefont {J.~N.}\ \bibnamefont
  {Reddy}},\ }\href@noop {} {\emph {\bibinfo {title} {Energy principles and
  variational methods in applied mechanics}}}\ (\bibinfo  {publisher} {John
  Wiley \& Sons},\ \bibinfo {year} {2017})\BibitemShut {NoStop}%
\bibitem [{\citenamefont {G{\'e}radin}\ and\ \citenamefont
  {Rixen}(2014)}]{geradin2014mechanical}%
  \BibitemOpen
  \bibfield  {author} {\bibinfo {author} {\bibfnamefont {M.}~\bibnamefont
  {G{\'e}radin}}\ and\ \bibinfo {author} {\bibfnamefont {D.~J.}\ \bibnamefont
  {Rixen}},\ }\href@noop {} {\emph {\bibinfo {title} {Mechanical vibrations:
  theory and application to structural dynamics}}}\ (\bibinfo  {publisher}
  {John Wiley \& Sons},\ \bibinfo {year} {2014})\BibitemShut {NoStop}%
\end{thebibliography}%

\end{document}